# Three-dimensional topological solitons in $\mathcal{PT}$-symmetric optical lattices

Yaroslav V. Kartashov,[1,2,*] Chao Hang,[3] Guoxiang Huang,[3] and Lluis Torner[1,4]

[1]*ICFO-Institut de Ciencies Fotoniques, The Barcelona Institute of Science and Technology, 08860 Castelldefels (Barcelona), Spain*
[2]*Institute of Spectroscopy, Russian Academy of Sciences, Troitsk, Moscow Region, 142190, Russia*
[3]*State Key Laboratory of Precision Spectroscopy and Department of Physics, East China Normal University, and NYU-ECNU Joint Institute of Physics at NYU-Shanghai, Shanghai 200062, China*
[4]*Universitat Politecnica de Catalunya, 08034, Barcelona, Spain*
*\*Corresponding author: Yaroslav.Kartashov@icfo.eu*

**We address the properties of fully three-dimensional solitons in complex parity-time ($\mathcal{PT}$)-symmetric periodic lattices with focusing Kerr nonlinearity, and uncover that such lattices can stabilize both, fundamental and vortex-carrying soliton states. The imaginary part of the lattice induces internal currents in the solitons that strongly affect their domains of existence and stability. The domain of stability for fundamental solitons can extend nearly up to the $\mathcal{PT}$-symmetry breaking point, where the linear lattice spectrum becomes complex. Vortex solitons feature spatially asymmetric profiles in the $\mathcal{PT}$-symmetric lattices, but they are found to still exist as stable states within narrow regions. Our results provide the first example of continuous families of stable three-dimensional propagating solitons supported by complex potentials.**

Generation of three-dimensional (3D) solitons has been a salient problem of fundamental importance since the birth of nonlinear physics. The problem critically hinges in the elucidation of physical settings that allow existence of three-dimensional self-sustained excitations that, in addition, are stable upon propagation. The latter requirement is particularly challenging because the common cubic (Kerr) nonlinearity present in most potentially suitable materials leads to supercritical collapse and thus cannot support stable higher-dimensional solitons in uniform media [1,2]. A number of approaches to stabilize 3D solitons have been suggested over the years [3,4]. Most of them suggest using nonlinearities that are different from pure cubic nonlinearity, introduce various higher-order effects, or rely on spatial modulations of system parameters. Thus, it has been predicted that stable 3D solitons (termed *light bullets* in optics) can be supported by media with saturable [5] and quadratic [6-8] nonlinearities, materials with competing [9-11] or nonlocal [12-15] nonlinearities, off-resonant two-level systems [16], etc. Stable evolution of 3D solitons is shown to be possible under the action of higher-order effects arising upon filamentation [17] or in the presence of higher-order dispersion or nonparaxial corrections [18]. 3D solitons were studied in dissipative settings too, where in addition to competing conservative nonlinearities, higher-order absorption is usually present [19-25].

A powerful strategy for realization of stable 3D bullets (as well as of topological 2D states [26,27]) relies on the spatial modulations of material parameters. Thus, longitudinal tandems [28,29], graded-index fibers [30], and the structures with combined transverse and longitudinal modulations [31], such as shaken optical lattices [32], can be used to suppress collapse. Stabilization of 3D solitons in a transversally *periodic* medium was predicted in discrete waveguide arrays [33-35]. This idea was transferred to continuous optical systems [36-39] and also to Bose-Einstein condensates (BECs) [40-42]. Importantly, this strategy afforded the first experimental observation of fundamental light bullets in fiber arrays [43] that was later extended to vortex solitons [44]. In all previous works, the transverse linear potential stabilizing the 3D solitons was real, i.e. it was created only by the refractive index modulation (in optics) or by using standing optical lattices (in BECs). Nevertheless, the recent discovery [45] of a new class of so-called $\mathcal{PT}$-symmetric complex potentials, obeying the condition $\mathcal{R}(\mathbf{r}) = \mathcal{R}^*(-\mathbf{r})$ and possessing purely real spectra when $\text{Im}\,\mathcal{R}(\mathbf{r})$ is below certain critical level, has opened important new horizons for the control of linear and nonlinear light propagation.

The concept of $\mathcal{PT}$-symmetry introduced initially in quantum mechanics upon construction of non-Hermitian operators with real spectra has been extended to several areas of science. Among them are lasers with single-mode operation and enhanced tunability [46,47], resonant systems such as atomic gases [48], design of metamaterials and metasurfaces where symmetry breaking may manifest itself in changes of the state of polarization of reflected light [49], formation of light-matter excitation in dissipative exciton-polariton condensates [50], micro-cavities [51], or acoustic devices [52], to name a few. Importantly, many of these systems may operate in the nonlinear regime where under appropriate conditions self-sustained nonlinear states may form. Elucidation of suitable conditions for the existence of such stable excitations, especially multidimensional ones, is a topic of continuously renewed interest. Optics provides the ideal setting to study the concept.

It should be mentioned that optical waveguiding systems offer particularly convenient platforms for the realization of $\mathcal{PT}$-potentials [53,54], because they allow the simultaneous modulation of the refractive index and the gain and losses. $\mathcal{PT}$-symmetry has been experimentally observed in several settings [55,56]. There is considerable interest in the investigation of solitons in $\mathcal{PT}$-symmetric potentials. The properties of 1D and 2D solitons in periodic $\mathcal{PT}$-symmetric lattices are known [57-66], but 3D solitons have not been achieved. Note that previous works [67,68] about 3D solitons used complex potentials localized in three dimensions and employed symmetries absent in a periodic medium. Thus, the question of whether stable 3D solitons are possible in complex potentials remains open. In this paper we show for the first time that stable fundamental and vortex 3D solitons can form in 2D $\mathcal{PT}$-symmetric periodic optical lattices imprinted in focusing cubic media. To the best of our knowledge, this is the first known example of stable 3D solitons supported by complex periodic lattices that do not require higher-order dissipation for stable propagation.

We address the propagation of a spatiotemporal wavepacket along the $\xi$-axis of a cubic nonlinear optical medium with anomalous dispersion and imprinted transverse modulation of the refractive index and gain/losses. The evolution of the wavepacket is governed by the dimensionless nonlinear Schrödinger equation:

$$i\frac{\partial q}{\partial \xi} = -\frac{1}{2}\left(\frac{\partial^2 q}{\partial \eta^2} + \frac{\partial^2 q}{\partial \zeta^2}\right) - \frac{\beta}{2}\frac{\partial^2 q}{\partial \tau^2} - q|q|^2 - \mathcal{R}(\eta,\zeta)q. \qquad (1)$$

Here $q = (2\pi L_{\text{dif}} n_2/\lambda)^{1/2} E$ is the dimensionless field amplitude; $n_2$ is the nonlinear coefficient; $L_{\text{dif}} = k r_0^2$ is the diffraction length; $k = 2\pi n/\lambda$ is the wavenumber; $n$ is the refractive index; $(\eta,\zeta) = (x r_0^{-1}, y r_0^{-1})$ and $\xi = z/L_{\text{dif}}$ are transverse and longitudinal coordinates normalized to the characteristic transverse scale $r_0$ and diffraction length $L_{\text{dif}}$, respectively; $\tau = t/t_0$ is the normalized (retarded) time; $\beta = L_{\text{dif}}/L_{\text{dis}}$ (further we set $\beta = 1$); $L_{\text{dis}} = t_0^2/|\partial^2 k/\partial \omega^2|$ is the dispersion length. The function $\mathcal{R} = p_{\text{r}}[\cos(\Omega \eta) + \cos(\Omega \zeta)] + i p_{\text{i}}[\sin(\Omega \eta) + \sin(\Omega \zeta)]$ describes the profile of complex lattice, where $p_{\text{r,i}} = \delta n_{\text{r,i}} k^2 r_0^2/n$ values are determined by the actual modulation depth of the complex refractive index $\delta n_{\text{r}} + i\delta n_{\text{i}}$, and $\Omega = 4$ is the lattice frequency. This complex 2D lattice is $\mathcal{PT}$-symmetric, i.e. $\mathcal{R}(\eta,\zeta) = \mathcal{R}^*(-\eta,-\zeta)$. Here we assume that the lattice profile does not change in time. Note that temporal lattices have been created experimentally [69], but the extension of the technique to realize the model studied here is not straightforward. In contrast, since doping of fibers with active or absorbing centers is a well-established technology, properly designed arrays of dissipative fibers, similar to conservative arrays from [43,44], may be potentially readily used for the experimental observation of the phenomena described here. Indeed, spatial modulation of gain and losses may be realized by packing absorbing and amplifying fibers into one complex periodic array and using properly shaped pump beams. By assuming similar nonlinear and dispersive properties of the material as in the experiments that addressed the formation of conservative light bullets that were performed in waveguide arrays with silica cores [43,44], for a characteristic transverse scale of $r_0 = 50\ \mu\text{m}$, the diffraction length at $\lambda = 1550$ nm ($n \approx 1.45$) amounts to about $L_{\text{dif}} \sim 14.6$ mm, the frequency $\Omega = 4$ corresponds to a lattice period of some $78\ \mu\text{m}$, and the dispersion coefficient $-280\ \text{fs}^2/\text{cm}$ yields a characteristic temporal scale of the order of $t_0 \sim 20$ fs. Under such conditions, a depth $p_{\text{r}} = 6$ of the real part of the lattice corresponds to a refractive index contrast of the order of $\delta n_{\text{r}} \sim 10^{-4}$, while a depth $p_{\text{i}} = 1$ of its imaginary part corresponds to a gain/loss amplitude of about $\sim 0.7$ cm$^{-1}$. Assuming a nonlinear coefficient of $n_2 = 2.2 \times 10^{-16}$ cm$^2$/W and a Gaussian-like beam profile a dimensionless intensity $|q|^2 = 1$ corresponds to some $\sim 76$ GW/cm$^2$. Experiments with pulsed excitation in $\mathcal{PT}$-symmetric systems may also be performed with other materials featuring much stronger nonlinearities, see e.g. [55,56].

On rigorous grounds, once dispersion of the dielectric permittivity of the medium is taken into account, the exact $\mathcal{PT}$-symmetry may be realized only for a discrete set of optical frequencies, as follows from the Kramers-Kronig relations [70]. Nevertheless, as long as gain and losses are provided by dopants in such low concentrations that do not notably affect the dispersion of the real part of the permittivity of the host material, as it is assumed here, and, in turn, as long as the imaginary part of the permittivity is determined exclusively by the dopants, the deviations from the exact $\mathcal{PT}$-symmetric conditions may not have a significant effect within the frequency range occupied by the propagating soliton pulses. Moreover, gain and losses may be provided by different mechanisms, including nonlinear ones [56], which may further reduce or even vanish the impact of the issue.

As a specific setting for the experimental implementation of the model (1) we also suggest $\mathcal{PT}$-symmetric refractive index landscapes imprinted in a cold gas of two atomic isotopes in a Λ-type configuration (e.g., $^{87}$Rb and $^{85}$Rb isotopes) loaded in an atomic cell, as described in [48]. Due to the interference of two Raman resonances, the required spatial distribution of the refractive index may potentially be realized by using a proper combination of a control laser field and a far-off-resonance Stark laser field. Since the refractive index of the atomic vapor is determined by the two external laser fields whose intensities could be at microwatt levels, one may use them for fine-tuning the $\mathcal{PT}$-symmetric potential. In addition, such a system is characterized by a large Kerr nonlinearity (sometimes exceeding usual non-resonant nonlinearities by many orders of magnitude [71]) that may favor the formation of solitons at remarkably low power levels. At the same time, inducing the required dispersion in such atomic system may be a challenge.

The existence domains of 3D solitons are closely connected to the existence domains of *linear spatial* Bloch modes $a_m(\eta,\zeta)\exp(ib_m\xi + ik_\eta \eta + ik_\zeta \zeta)$ of the 2D lattice $\mathcal{R}(\eta,\zeta)$, where $b_m = b_m^{\text{re}} + ib_m^{\text{im}}$ is the propagation constant, which can be complex, $m$ is the band index, $2\pi/\Omega$-periodic function $a_m(\eta,\zeta)$ describes the Bloch wave shape with momentum $(k_\eta, k_\zeta)$. Substitution of the field in such a form into the linear version of Eq. (1) without the temporal derivative yields the linear eigenvalue problem:

$$b_m a_m = \frac{1}{2}\left(\frac{\partial^2}{\partial \eta^2} + \frac{\partial^2}{\partial \zeta^2} + 2ik_\eta \frac{\partial}{\partial \eta} + 2ik_\zeta \frac{\partial}{\partial \zeta} - k_\eta^2 - k_\zeta^2\right)a_m + \mathcal{R}(\eta,\zeta)a_m, \qquad (2)$$

which we solved by a plane-wave expansion method. Due to the periodicity of $\mathcal{R}(\eta,\zeta)$ all allowed propagation constant values are arranged into bands. For example, for $p_{\text{r}} = 6$ and $p_{\text{i}} = 0$ (i.e. for real potential) two upper bands are separated by a forbidden gap. The dependencies $b_m(k_\eta, k_\zeta)$ for nonzero $p_{\text{i}}$ values associated with first two bands are shown in Figs. 1(a),(b) for complex potentials with $p_{\text{i}} < p_{\text{r}}$ and $p_{\text{i}} > p_{\text{r}}$, respectively. For $p_{\text{i}}$ values below the $\mathcal{PT}$-symmetry breaking point $p_{\text{i}} = p_{\text{r}}$ the spectrum remains entirely real, i.e. $b_m^{\text{im}} \equiv 0$, and first and second bands remain separated for all $(k_\eta, k_\zeta)$. At $p_{\text{i}} = p_{\text{r}}$ these bands touch each other along the entire periphery of the first Brillouin zone. Increasing $p_{\text{i}}$ beyond the critical value $p_{\text{i}} = p_{\text{r}}$ causes merging of the real parts of $b_{1,2}$ in some fraction of the first Brillouin zone and appearance of nonzero imaginary parts $b_m^{\text{im}}$ taking the largest values at $(k_\eta, k_\zeta) = (\pm \Omega/2, \pm \Omega/2)$ [Fig. 1(c)]. The maximal real part $b_{\text{re}}^{\text{max}}$ of the propagation constant defining the lower edge of the semi-infinite forbidden gap is achieved at $(k_\eta, k_\zeta) = (0,0)$ and it crosses zero exactly in the point where the lattice spectrum becomes complex [Fig. 1(c), solid line]. Figure 1 reveals two conditions for the existence of stable 3D bright solitons. First, solitons can only be stable if the background located far from soliton center is stable. This means that stable nonlinear modes can be found only in the regime with unbroken $\mathcal{PT}$-symmetry, at $p_{\text{i}} < p_{\text{r}}$. Second, in a nonlinear dispersive medium, exponential localization of the tails of solitons is guaranteed only when $b$ remains above the edge of the continuous spectrum both in the temporal and spatial domains. Thus, the lower cutoff for the existence of the 3D solitons is given by $b_{\text{co}} = \max[b_{\text{re}}^{\text{max}}, 0]$.

Figure 2 summarizes the main properties of the stationary 3D solitons supported by $\mathcal{PT}$-symmetric lattices, and Fig. 3 illustrates representa-tive profiles of such states. Stationary soliton solutions were obtained from Eq. (1) in the form $q=a(\eta,\zeta,\tau)\exp(ib\xi)$, where the function $a$ describes the soliton shape that remains unchanged upon propagation in the absence of perturbations. Here $a$ is complex, since even for a fundamental solution the internal energy currents from amplifying to absorbing domains appear due to nonzero $\mathrm{Im}\,\mathcal{R}(\eta,\zeta)$, but the propagation constant $b$ is always real, since we are searching for nonlinear states for which gain and losses are integrally balanced. Solutions were obtained with the method of squared operator described in Ref. [72] using fast Fourier transforms similarly to spectral renormalization methods [73]. The method consists in the iterative calculation of the soliton shape using the expression $a^{(n+1)}=a^{(n)}-[\mathcal{N}^{-1}\mathcal{L}_1^{\dagger}\mathcal{N}^{-1}\mathcal{L}_0 a^{(n)}]d\xi$, where $\mathcal{L}_0 a=-(1/2)\Delta a+|a|^2 a-\mathcal{R}a+ba$. For exact soliton solutions $\mathcal{L}_0 a\equiv 0$. $\Delta=\partial^2/\partial\eta^2+\partial^2/\partial\zeta^2+\partial^2/\partial\tau^2$ is the Laplacian; the operator $\mathcal{L}_1(\delta a)=\mathcal{L}_0(a+\delta a)$ is obtained by linearizing $\mathcal{L}_0$ assuming small perturbations $\delta a\ll a$. Finally, the operator $\mathcal{N}a=(c-\Delta)a$ ensures convergence of the iterative procedure, with $c$ being a free parameter determining the convergence speed.

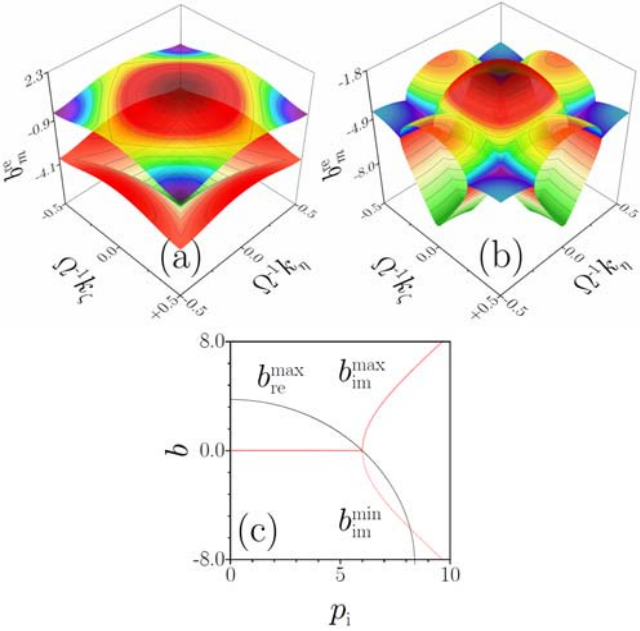

Fig. 1. (Color online) Band-gap spectrum of the two-dimensional $\mathcal{PT}$-symmetric lattice below [panel (a), $p_i=5$] and above [panel (b), $p_i=7$] the $\mathcal{PT}$-symmetry breaking point. Only the real part of the propagation constant in the first two bands is shown. (c) Evolution of maximal real $b_{\mathrm{re}}^{\max}$ and maximal and minimal imaginary $b_{\mathrm{im}}^{\max}, b_{\mathrm{im}}^{\min}$ parts of propagation constants of Bloch waves from the first two bands with an increase of $p_i$. In all cases, $p_r=6$.

Eq. (1) is intrinsically dissipative, therefore the total energy $U=\iiint |q|^2\,d\eta d\zeta d\tau$ is not a rigorous conserved quantity anymore (there exist inputs for which $U$ grows or decays upon propagation). Instead, conservation of the quantity $P=\iiint q(\eta,\zeta,\tau)q^*(-\eta,-\zeta,\tau)d\eta d\zeta d\tau$ is guaranteed by the $\mathcal{PT}$-symmetry of $\mathcal{R}(\eta,\zeta)$. Nevertheless, it is convenient to use the $U(b)$ dependencies to characterize the soliton families, since the energy $U$ is a directly measurable experimental parameter and it does not change upon evolution when a stationary unperturbed state is considered. Note that the existence of entire soliton families parameterized by the propagation constant $b$ despite the presence of gain and losses is a unique property of $\mathcal{PT}$-symmetric potentials.

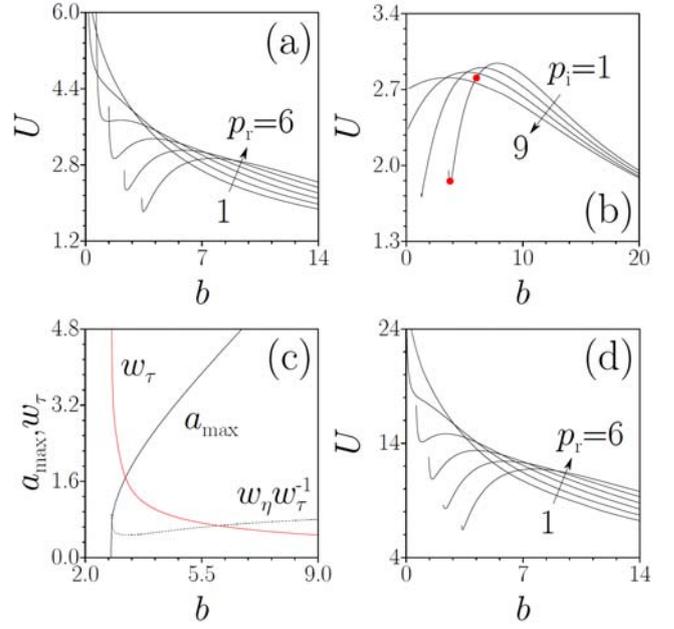

Fig 2. (Color online) Dependencies of energy of a fundamental 3D soliton on $b$ (a) for different $p_r$ values at $p_i=2$ and (b) for different $p_i$ values at $p_r=6$. In (a) the depth of the real part of the lattice increases by 1 in the direction shown by the arrow. In (b) the curves are shown for $p_i=1,5,7$ and $9$. (c) Peak amplitude $a_{\max}$ of fundamental light bullet, its temporal width $w_\tau$, and the ratio of spatial and temporal widths $w_\eta w_\tau^{-1}$ as functions of $b$ for $p_r=6$, $p_i=3$. (d) It is the same as in (a), but for the 3D vortex solitons.

Figure 2(a) illustrates how a monotonically decreasing $U(b)$ dependence (for fundamental solitons) in a lattice with a shallow real part transforms into a strongly nonmonotonic dependence in a lattice with deep real part at $p_i=2$. Although the Vakhitov-Kolokolov stability criterion is not applicable directly to solitons in $\mathcal{PT}$-symmetric systems (because of the potential occurrence of drift instabilities) it is commonly observed that in complex lattices of this type only solutions from $dU/db>0$ branches can be potentially stable in focusing media. Thus, the real part of the lattice has to be deep enough to give rise to stable solitons [in Fig. 2, $p_r>3$ is required for appearance of nonmonotonic $U(b)$ curves for $p_i=2$]. As shown in the plot, we found that there is no energy threshold for the existence of fundamental 3D solitons in the $\mathcal{PT}$-symmetric lattice (the energy tends to zero when $b\to\infty$). Indeed, when solitons become strongly localized at large $b$ values, the lattice affects their shapes only weakly and the $U(b)$ curves for different $p_{r,i}$ values approach the monotonically decreasing dependence $U\sim(C/b)^{1/2}$, with $C\approx 44.3$, as in uniform medium [10,11]. Scaling arguments that can be used to obtain the latter dependence nevertheless become inapplicable when the lattice strongly affects soliton shapes, hence at intermediate $b$ values one observes considerable deviations from the $U\sim(C/b)^{1/2}$ law. Importantly, note that Fig. 2(a) shows that stability can potentially be achieved only within a finite interval of energies and propagation constants.

When propagation constant approaches the cutoff value $b_{\mathrm{co}}$, the soliton energy increases rapidly. In the regime $p_i<p_r$ with unbroken $\mathcal{PT}$-symmetry, this cutoff coincides with the upper edge of the first band $b_{\mathrm{re}}^{\max}$ in the spectrum from Fig. 1. Hence, close-to-cutoff solitons strongly expand both in space and time (see representative shapes in Fig. 3, left column), featuring multiple spatial oscillations. Increasing $b$ results in a higher localization of the solitons and in the gradual equilibration of its spatial and temporal widths (Fig. 3, central column). Figure 3 also reveals a nontrivial phase distributions $\psi$ even for fundamental solitons that are due to the $\mathcal{PT}$-symmetric lattice. Such distributions indicate on the existence of transverse energy currents $\mathbf{j}=|a|^2\nabla\psi$ [here

$\nabla = (\partial/\partial\eta, \partial/\partial\zeta, \partial/\partial\tau)$] that are especially pronounced between amplifying to absorbing domains, and that become stronger with the increase of the depth of the imaginary lattice part $p_i$. Indeed, by writing the soliton solutions in the form $q = |a|\exp(i\psi + ib\xi)$, one can derive the following system of equations for the field modulus and the current:

$$b|a| = (1/2)\nabla^2|a| - \mathbf{j}^2/(2|a|^3) + |a|\mathcal{R}_r + |a|^3, \qquad (3)$$
$$\nabla \cdot \mathbf{j} = 2|a|^2 \mathcal{R}_i,$$

where $\mathcal{R}_r = \mathrm{Re}\,\mathcal{R}$ and $\mathcal{R}_i = \mathrm{Im}\,\mathcal{R}$. The magnitude of the current is directly determined by the imaginary part of the potential. Moreover, the current $\mathbf{j}$ strongly impacts the soliton shape. It should be noted that the soliton phase varies not only in space, but also in time $\tau$. This is illustrated in the last row of Fig. 3. However, the temporal variation of the phase is rather slow in comparison with its spatial variation, so that the corresponding temporal current component $j_\tau = |a|^2 \partial\psi/\partial\tau$ is small and therefore affects only weakly the soliton shape in comparison with $j_{\eta,\zeta}$.

Figure 2(b) illustrates the impact of the *imaginary part* of the lattice on the properties of the 3D fundamental soliton families. Increasing $p_i$ results in a decrease of the cutoff for soliton existence. The cutoff vanishes for $p_i \geq p_r$. When the $\mathcal{PT}$-symmetry is unbroken, both spatial $w_\eta$ and temporal $w_\tau$ integral soliton widths, defined according to

$$w_\eta^2 = U^{-1} \iiint \eta^2 |a|^2\, d\eta d\zeta d\tau, \qquad (4)$$
$$w_\tau^2 = U^{-1} \iiint \tau^2 |a|^2\, d\eta d\zeta d\tau,$$

diverge at $b \to b_{co}$ where peak soliton amplitude $a_{max} = \max|a|$ vanishes [Fig. 2(c)]. The ratio of spatial and temporal widths $w_\eta/w_\tau$ approaches 1 for large $b$ values. For $p_i > p_r$ the peak amplitude remains finite at $b \to b_{co} = 0$ and delocalization manifests itself only in the temporal domain, where solitons acquire long tails. Increasing imaginary part of the lattice results in a gradual shift of the maximum of the $U(b)$ dependence toward $b = 0$, but the width of the domain with a positive slope may become even larger at $p_i \to p_r$ than at $p_i \to 0$ [Fig. 2(b)].

A qualitatively similar picture was obtained for 3D solitons carrying a nested off-site topological dislocation, or vortex light bullets. Such solitons are composed of four optical peaks located in the vicinity of four neighboring lattice maxima. Traces of a canonical, staircase vortex phase distribution and phase singularity are visible in Fig. 3 (right), but the entire phase distribution $\psi$ is rather complex due to the presence of internal currents from amplifying to absorbing domains. Snapshots of the intensity distributions carried by the spatio-temporal vortex solitons at different $\tau$ values reveal their spatial asymmetry. On physical grounds, such asymmetry is a consequence of the interference of the global current associated with vorticity and the local currents arising inside waveguides due to inhomogeneous gain and losses. Namely, in some waveguides the directions of global and local currents coincide, leading to an increase of total current, but in other waveguides the currents take opposite directions, which results in current reduction. As a consequence, the $|\mathbf{j}(\eta,\zeta)|$ distributions at any $\tau$ do not feature a four-fold rotation symmetry as in conservative lattices, and such spatial asymmetry directly translates into an asymmetry of the field distribution $|q|$ as it follows from Eq. (3). The effect is readily visible in Fig. 3, where we compare the field distributions (modulus) of the 3D spatio-temporal vortex solitons with positive and negative topological charges $l$. The asymmetry in the shape of the vortex solitons increases when the imaginary part of the lattice grows (compare $\tau = 0$ cross-sections for vortex solutions shown in right column of Fig. 3, obtained for $p_i = 6$, and the solution shown in Fig. 4, obtained for $p_i = 0.55$). Typical $U(b)$ curves for vortex solitons are presented in Fig. 2(d). Curves are qualitatively similar to those featured by fundamental solitons, but the energy carried by a vortex solitons is approximately four times higher than that of a fundamental soliton. It should also be noted that the 3D vortex solitons do not bifurcate from the upper edge of the band, i.e. their cutoff is slightly higher than that of fundamental solitons.

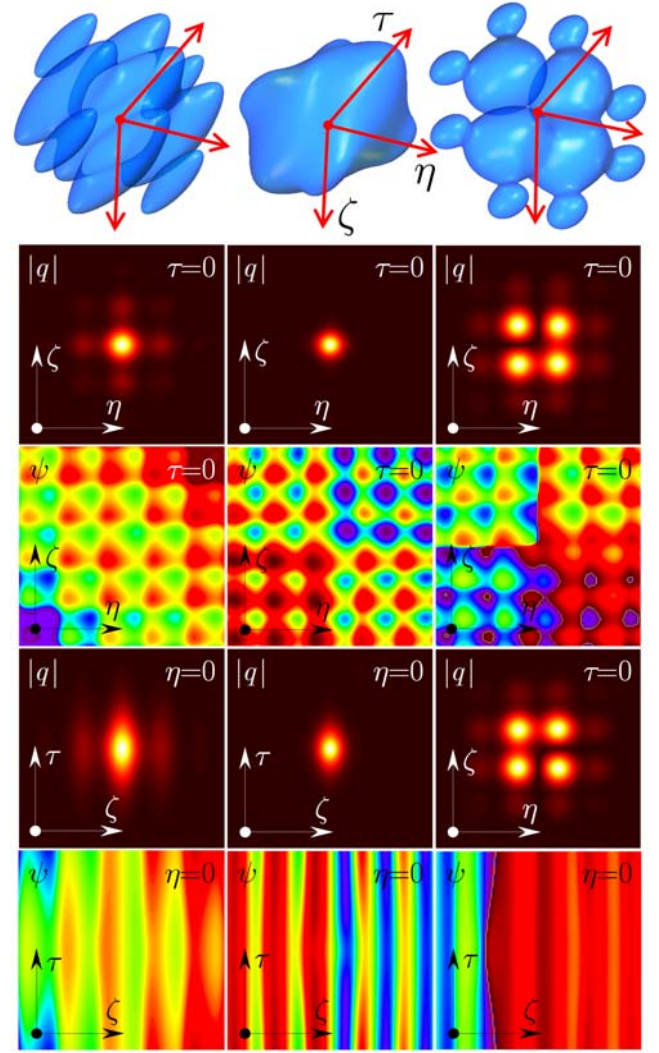

Fig. 3. (Color online) Profiles of stationary fundamental 3D solitons at $b = 3.8$, $p_i = 1$ (left column), $b = 6$, $p_i = 1$ (central column), and of $l = +1$ vortex solitons at $b = 0.5$, $p_i = 6$ (right column). The first row shows isosurface field modulus distributions at the level $|q| = 0.03$ (left and central columns) and at the level $|q| = 0.1$ (right column). The second and third rows show field modulus and phase distributions at $\tau = 0$, fourth row shows $|q|$ at $\eta = 0$ (except for the right column, where the fourth row shows $|q|_{\tau=0}$ for $l = -1$ soliton), and the fifth row shows phase $\psi$ at $\eta = 0$. Solitons in the left and central columns correspond to the red dots in Fig. 2(b). In all cases $p_r = 6$. Distributions in the second-fifth rows are shown in the $\eta, \zeta, \tau \in [-4, +4]$ windows.

The central result of this paper is summarized in Fig. 5, which depicts the stability properties of both, fundamental and vortex 3D solitons in $\mathcal{PT}$-symmetric lattices. In order to check stability we first resort to direct integration of Eq. (1) with inputs in the form of solitons solutions perturbed with weak small-scale noise. Such perturbed states were allowed to propagate for large distances, up to $\xi = 10^3$, which allow to identify the parameter domains where 3D solitons are dynamically stable. This method of stability analysis tests the presence of a wide range of potential instabilities, because the broadband input noise excites a correspondingly broad spectrum of perturbations. Due to the large propagation distances considered, exponentially growing modes are readily detectable. At the same time, because noise is introduced with small amplitude, the perturbation introduces only weak deformations into the input soliton that thus do not lead to pronounced longitudinal oscillations of its amplitude/width. We found that fundamental solitons are stable in a considerable part of their existence domain on the $(p_i, b)$ plane, nearly up to the $\mathcal{PT}$-symmetry breaking point [in the region between red dots in Fig. 5(a)]. The Vakhitov-Kolokolov criterion correctly predicts the lower border $b_{cr}^{low}$ of the stability domain for the fundamental solitons that is very close to the cutoff for soliton existence, but it does not predict its upper border $b_{cr}^{upp}$ for $p_i > 0$. This is because at $b = b_{cr}^{upp}$ oscillatory rather than exponential instabilities come into play. These are accompanied by transverse shift/oscillations of the soliton center. The stability domain shrinks at first with the increase of $p_i$, but then it expands again. It completely disappears slightly below the symmetry breaking point $p_i = p_r$, where the spatial background becomes unstable. We also found that the $\mathcal{PT}$-symmetric lattices can stabilize 3D vortex solitons, even though they are *spatially asymmetric*. The domain of stability for the vortex solitons [Fig. 5(b)] is much narrower than that of fundamental solitons in terms of $p_i$, but it is more extended in terms of $b$. The 3D vortex solitons are also be stable within the interval $b_{cr}^{low} < b < b_{cr}^{upp}$ of propagation constants. The domain of stability shrinks completely at $p_i \approx 0.6$ for $p_r = 6$, but its width increases with growing $p_r$. The development of instability for the 3D vortex solitons at $b > b_{cr}^{upp}$ is initiated by the concentration of energy in one of the constituent sub-peaks and its subsequent collapse. In contrast, fundamental 3D solitons with $b > b_{cr}^{upp}$ oscillate and radiate energy away until their parameters reach a stability interval and a new stable state is formed.

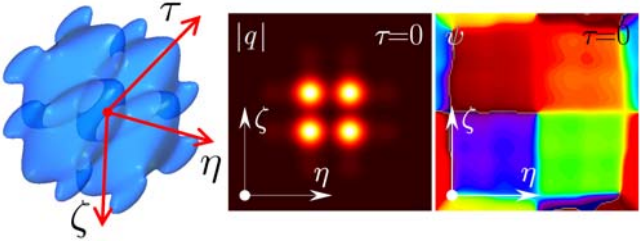

Fig. 4. (Color online) Profile of a stable $l = +1$ vortex soliton with $b = 4.4$, $p_i = 0.55$. Isosurface depicting the field modulus distribution (left) at $|q| = 0.07$.

To cross-check the above-mentioned results we performed also a linear stability analysis. Perturbed solutions were written in the form $q = [a + (u+v)e^{\delta\xi} + (u^* - v^*)e^{\delta^*z}]e^{ib\xi}$, where $u, v \ll a$ are complex functions describing the perturbation profile and $\delta$ is the perturbation growth rate. Substitution into Eq. (1) and its linearization yields the linear eigenvalue problem $\mathcal{L}\Psi = \delta\Psi$, where $\Psi = (u,v)^T$, with the matrix operator

$$\mathcal{L} = i \begin{pmatrix} \frac{\mathcal{A}_-}{2} + i\mathcal{R}_i & \frac{\Delta}{2} + 2|a|^2 - \frac{\mathcal{A}_+}{2} + \mathcal{R}_r - b \\ \frac{\Delta}{2} + 2|a|^2 + \frac{\mathcal{A}_+}{2} + \mathcal{R}_r - b & -\frac{\mathcal{A}_-}{2} + i\mathcal{R}_i \end{pmatrix} \quad (5)$$

where $\mathcal{A}_- = a^2 - a^{*2}$, $\mathcal{A}_+ = a^2 + a^{*2}$. Standard eigenvalue solvers can give solution of this 3D problem only for small number of transverse points (even though the corresponding matrix is sparse), a procedure that however does not guarantee sufficient accuracy. Instead, here we used the method put forward in [74] and obtained the eigenvalues and eigenfunctions of $\mathcal{L}$ by solving the equation $\partial\Psi/\partial\xi = \mathcal{N}^{-1}(\mathcal{L} - \delta)\Psi$, where we introduced the operator $\mathcal{N} = (c - \Delta)\operatorname{diag}(1,1)$ accelerating convergence that is analogous to the operator used to find soliton solutions above. This equation is solved using a fourth-order Runge-Kutta method. At each step in $\xi$ the eigenvalue $\delta$ is computed from $\Psi$ as $\delta = \langle \mathcal{N}^{-1}\Psi, \mathcal{L}\Psi \rangle / \langle \mathcal{N}^{-1}\Psi, \Psi \rangle$, where $\langle \Psi_1, \Psi_2 \rangle = \iiint \Psi_1^\dagger \Psi_2 d\eta d\zeta d\tau$ is a standard inner product. It was shown in Ref. [74] that for sufficiently broad class of input conditions, at large $\xi$ this method converges to the perturbation mode $\Psi$ whose eigenvalue $\delta$ has the largest real part $\delta_r^{max}$, provided that the corresponding solution is dynamically unstable. A typical dependence $\delta_r^{max}(b)$ calculated with the described approach is shown in Fig. 5(c) for the case of 3D vortex solutions. The plot confirms that linear stability analysis predicts stable vortex soliton solutions (i.e., with $\delta_r^{max} = 0$) in the same interval of propagation constants that was obtained by direct propagation of the perturbed solutions presented above. The typical normalized decay distance for unstable soliton solutions can be estimated as $\xi \sim 1/\delta_r^{max}$, which corresponds to an actual length of $L_{dif}/\delta_r^{max}$.

Summarizing, it is readily apparent that inclusion of the dissipative part of the potential may drastically modify the properties of the system. As it becomes non-Hamiltonian, strong internal currents emerge that substantially modify soliton shapes that are especially pronounced in the case of vortex solitons, and instabilities come into play that may destroy even fundamental solitons in the regions where they are otherwise stable in conventional conservative systems.

Nevertheless, we have revealed that complex (i.e., with real and imaginary parts arising from gain and losses) $\mathcal{PT}$-symmetric two-dimensional lattices are capable of supporting robust, stable fully three-dimensional fundamental and vortex solitons even in Kerr nonlinear media. We showed that the strength of gain/losses acting in the system affects strongly the soliton parameters and the corresponding stability domains. The states discovered here in the $\mathcal{PT}$-symmetric systems are the first known examples of stable, three-dimensional, *propagating soliton families* supported by complex potentials. It is readily apparent that dissipative $\mathcal{PT}$-symmetric lattices offer additional tools for controlling the shapes, domains of existence, and stability domains of three-dimensional excitations as compared to conservative lattices. In particular, the dissipative effects introduce features in the soliton shapes and stability properties that are not present in comparable conservative settings.

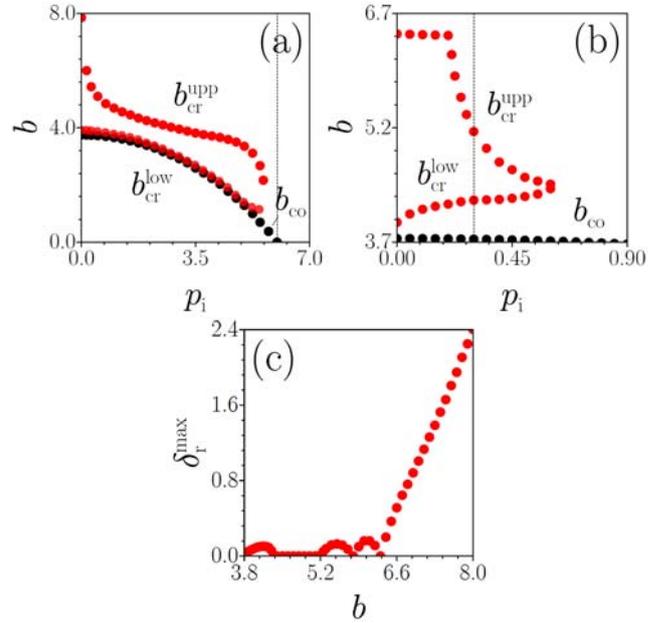

Fig. 5. (Color online) Domains of stability on the plane $(p_i, b)$ for fundamental (a) and vortex (b) 3D solitons at $p_r = 6$. Solitons are stable in the region between the red dots at $b_{cr}^{low} < b < b_{cr}^{upp}$. The black dots denote the cutoff for the fundamental 3D solitons. (c) Real part of perturbation growth rate versus $b$ for vortex 3D solutions for $p_i = 0.3$, corresponding to the vertical dashed line in panel (b).

**Acknowledgement:** This work is partially supported by the Severo Ochoa Excellence program of the Government of Spain and by Fundació Cellex (Barcelona). The Chinese co-authors were supported by the Pro-gram of Introducing Talents of Discipline to Universities (B12024), and by the NSF-China (grants No.~11475063 and No.~11474099).